\documentclass[11pt]{article}

\usepackage[iso-8859-7]{inputenc}
\usepackage{epsfig}
\usepackage{graphicx}
\usepackage{epstopdf}
\usepackage{amsfonts}
\usepackage{amssymb}
\usepackage{amsthm}
\usepackage{amsmath}
\usepackage{multirow}

\newcommand{\newc}{\newcommand}
\newc{\ra}{\rightarrow}
\newc{\lra}{\leftrightarrow}
\newc{\be}{\begin{equation}}
\newc{\ee}{\end{equation}}
\newc{\bs}{\begin{split}}
\newc{\es}{\end{split}}
\newc{\ba}{\begin{eqnarray}}
\newc{\ea}{\end{eqnarray}}
\newc{\ov}{\overline}
\newc{\pa}{\partial}
\newc{\D}{\Delta}
\def\eps{\epsilon}

\newc{\nn}{\nonumber}

\newcommand\myeqr{\mathrel{\stackrel{\makebox[0pt]{\mbox{\normalfont\tiny $Z_2$}}}{\longrightarrow}}}
\begin{document}
	\begin{titlepage}
		
		%\vspace*{-15mm}
		%\begin{flushright}
		%SHEP-11-XX\\
		%\end{flushright}
		\vspace*{0.7cm}

		\begin{center}
{\Large {\bf Phenomenology with F-theory $SU(5)$}}
			\\[12mm]
			George~K.~Leontaris$^{a,b}$
			\footnote{E-mail: \texttt{leonta@uoi.gr}} and
			Qaisar~Shafi$^{c}$
			\footnote{E-mail: \texttt{shafi@bartol.udel.edu}}
			\\[-2mm]
			
		\end{center}
		\vspace*{0.45cm}
		\centerline{$^{a}$ \it
			LPTHE, UMR CNRS 7589, Sorbonne Universit\'es,}
		\centerline{\it
		 UPMC Paris 6, 75005 Paris, France }
		\vspace*{0.15cm}
			\centerline{$^{b}$ \it
					Physics Department, Theory Division, University of Ioannina,}
				\centerline{\it
					GR-45110 Ioannina, Greece }
				\vspace*{0.2cm}
		\centerline{$^{c}$ \it
		Bartol Research Institute, Department of Physics and Astronomy, University of Delaware,}
		\centerline{\it
			DE 19716,  Newark, USA}
		\vspace*{1.20cm}
		
		\begin{abstract}
\noindent	
We explore the low energy phenomenology of an F-theory based SU(5) model which, in addition to the known quarks and leptons,
contains Standard Model (SM)  singlets, and vector-like color triplets and SU(2) doublets. Depending on their masses and couplings,
some of these new particles may be observed at the LHC and future colliders. We discuss the restrictions by CKM constraints on their 
mixing with the ordinary down quarks of the three chiral familes. The model is consistent with gauge coupling unification
at the usual supersymmetric GUT scale, dimension five proton decay is adequately suppressed, while dimension-six decay mediated by the
superheavy gauge bosons is enhanced by a factor of 5-7. The third generation charged fermion Yukawa couplings  yield the corresponding
 low-energy masses in reasonable agreement with observations. The hierarchical nature of the masses of lighter generations is accounted
for via non-renormalisable interactions,  with the perturbative vacuum expectation values (vevs) of the SM singlet fields playing an
essential r\^ole.

		\end{abstract}
		
	\end{titlepage}

 % % % % % % % % % % % % % % % % % % %

 \section{Introduction}

Models originating from string theory constructions often contain SM singlets and vector-like fields
which can mix with the light spectrum and therefore are natural candidates for predicting rare processes
 that might be discovered in  future  experiments at the LHC and elsewhere.  F-theory models~\cite{Beasley:2008dc},
in particular, have the necessary ingredients to describe in a simple and convincing manner a complete picture of
such new phenomena. One of the most appealing grand unified theories incorporating these features in an F-theory context,
is $SU(5)$~\footnote{For  F-theory model building reviews and early references
see~\cite{Weigand:2010wm,Heckman:2010bq,Leontaris:2012mh,Maharana:2012tu}.
For an incomplete list including more recent research papers
see~\cite{Donagi:2009ra}-\cite{Romao:2015jrh}.}. Indeed, on breaking  F-$SU(5)$  to SM symmetry, one ends up with the
MSSM spectrum augmented by scalar fields and vector-like states, which are remnants of the underlying GUT representations.
In this framework, it is possible to retain gauge coupling unification even in the presence of some additional fields,
provided that these form complete multiplets of $SU(5)$.  In view of the ongoing experimental searches and possible future
signatures, in this work we reconsider some issues regarding the exotic part of these models.

We start with a brief review of the basic features of an $SU(5)$ model~\cite{Leontaris:2010zd} derived in an F-theory framework
 and, in particular, in the context of the spectral cover.  We derive an effective  theory model by imposing a $Z_2$ monodromy
 and identify the complex surfaces where the chiral matter and Higgs can be accommodated in the quotient theory. We assume a
 hypercharge flux breaking of the $SU(5)$ symmetry down to the SM one, and proceed with a specific assignment of  the MSSM
  representations on these matter curves and then work out the spectrum and the superpotential.  After fixing the necessary free
parameters (such as flux units and singlet vevs), we proceed with the investigation of the exotic massless spectrum left over from
higher dimensional    fields.  We then derive their superpotential couplings and analyse the implications for baryon  number violating
decays as well as other rare processes. We examine the possibility that these states remain  massless at low energies
being consistent with gauge coupling  unification, and discuss the physics implications of the TeV scale exotic states.

\section{F-SU(5)}

We consider the elliptically fibred case where  the highest smooth singularity  in Kodaira's classification is associated with
the exceptional group of $E_8$~\cite{Vafa:1996xn,Morrison:1996na}. We assume 7-branes wrapping an $SU(5)$ divisor and interpret
 this as the GUT  symmetry of the effective model. Under these assumptions
\be
E_8\supset SU(5)_{GUT}\times  SU(5)_{\perp}\,,\label{su5}
\ee
where the first factor is interpreted as the well known $SU(5)_{GUT}$ and the second factor is usually denoted as $SU(5)_{\bot}$.

\noindent
The MSSM spectrum and possible exotic fields descend from the decomposition of the $E_8$ adjoint which, under
the assumed breaking pattern (\ref{su5}), decomposes as follows:
\ba
248&\ra& (24,1)+(1,24)+(10,5)+(\overline{5},10)+(5,\overline{10})+(\overline{10},\overline{5})\,.\label{E8To5}
\ea
Thus, matter transforms in bi-fundamental representations, with
the GUT  10-plets  lying in the fundamental of  $SU(5)_{\perp}$, and the
$\bar 5, 5$-plets lying in the antisymmetric representation of $SU(5)_{\perp}$.

We choose to work in the Higgs bundle picture  (the spectral cover approach).
In this context the properties of the GUT representations with respect to the
spectral cover are described by  a degree-five polynomial~\cite{Donagi:2009ra}
\be
\begin{split}
{\cal C}_5:&\qquad \sum_{k=0}^5 b_ks^{5-k}=0\,,\label{SC5}
\end{split}
\ee
where the $b_k$ coefficients carry the information of the internal geometry
and their homologies, are given by $[b_n]=\eta-nc_1$, (with  $\eta=6c_1-t$), where $c_1=c_1(S)$ is the first
Chern class of the tangent bundle and $-t$ that of the normal to the surface $S$.
The  roots of the equation are identified as the weight vectors $t_{1,...,5}$  satisfying the
standard $SU(N)$ constraint ($N=5$ in the present case)
\be
\sum_{i=1}^5t_i = 0\,.\label{sc5}
\ee
Under $t_i$  the  matter curves acquire specific topological and
symmetry properties inherited by the fermion families and Higgs fields propagating there.
We denote the matter curves accommodating the 10-plets, 5-plets  of $SU(5)$ and singlets
emerging from $SU(5)_{\perp}$ adjoint decomposition as
$\Sigma_{10_{t_i}}, \Sigma_{5_{t_i+t_j}},  \Sigma_{1_{t_i-t_j}}.$
Correspondingly, the possible representations residing on these matter curves are denoted by
\[ \Sigma_{10_{t_i}}: 10{_{t_i}}, \ov{10}{_{-t_i}},\;\;  \Sigma_{5_{t_i+t_j}}:\, \ov{5}_{t_i+t_j}\,,
 5_{-t_i-t_j},\;\; \Sigma_{1_{t_i-t_j}}:\, 1_{t_i-t_j}\,,  \]
where, as far as 5-plets and singlets are concerned, we must have $t_i\ne t_j$.

\noindent
Working in the framework of spectral cover, while assuming  distinct roots $t_i$ of~(\ref{SC5}),
one may further consider the breaking $SU(5)_{\bot}\to U(1)_{\bot}^4$.
Then, the invariant tree-level superpotential  couplings are of the form
\be
{\cal W}\supset
h_1\,10_{t_i}10_{t_j} 5_{-t_i-t_j}+h_2\,10_{t_i}\bar 5_{t_j+t_k}\bar 5_{t_l+t_m}+
h_3\, 1_{t_i-t_j} 5_{-t_i-t_k}\bar 5_{t_j+t_m} +h_4\, 1_{t_i-t_j} \, 1_{t_j-t_k} \, 1_{t_k-t_i},
\label{esup}
\ee
where $h_{1,2,3,4}$ represent the Yukawa strengths. In each of the above terms, the sum of the
$t_i$ `charges' should add up to zero. Hence, in the second term  ${t_i}+{t_j+t_k}+{t_l+t_m}=0$,
which unambiguously implies  that all indices in the term proportional to Yukawa coupling  $h_2$
should differ from each other (due to the fact that ${t_1}+{t_2+t_3}+{t_4+t_5}=0$).

Returning to the polynomial~(\ref{SC5}), although its coefficients $b_n$ belong to a certain field
(holomorphic functions), the roots $t_i$ do not necessarily do so. Solutions, in general, imply branch
cuts and, as a result,  certain roots might be interrelated.  The simplest case is if two of them are subject
to a $Z_2$ monodromy,  say,~\footnote{For various choices of monodromies,
	see~\cite{Bouchard:2009bu,Marsano:2009gv,Dudas:2010zb,Antoniadis:2012yk}.}
\be
Z_2:\, t_1=t_2\,.\label{Z2M}
\ee
From the point of view of the  effective field theory model, the appearance  of the monodromy is a welcome result
 since it implies rank-one mass matrices for the fermions. Indeed, under the  $Z_2$ monodromy, the coupling
 \be
 {\cal W}\supset 10_{t_1}10_{t_2} 5_{-t_1-t_2} \stackrel{Z_2}\longrightarrow  10_{t_1}10_{t_1} 5_{-2t_1}
\ee
ensures a top-quark mass at tree-level, while the remaining mass matrix entries are expected to be generated
from non-renormalisable terms.  After this brief description of the basic features,  we proceed in the next section
with the analysis of the implications of the hypercharge flux on the symmetry breaking  and the massless spectrum of $SU(5)$.

\section{Hypercharge Flux  breaking of $SU(5)$ }

The $Z_2$ monodromy implies that the spectral cover polynomial factorises as follows:
\be
b_0s^5+b_2s^3+b_3s^2+b_4s+b_5= ( a_1+a_2s+a_3s^2)(a_4+a_5s)(a_6+a_7s)(a_8+a_9s),\label{Z2split}
\ee
where all $a_i$ are assumed in the same field as $b_n$'s. Thus,  while the roots of
the three monomials on the right-hand side of~(\ref{Z2split}) are rational functions
in this field, it is assumed that the two roots of the binomial $( a_1+a_2s+a_3s^2)$ cannot be written
in terms of functions in the same field.

\noindent
The $b_n(a_i)$  relations are easily extracted  by identifying coefficients of the same powers in $s$
and are of the form $b_n=\sum a_ia_ja_ka_l$, where the indices satisfy $i+j+k+l+n=24$. Therefore, given
the homologies $[b_n]$, the corresponding  ones for the $a_i$ coefficients satisfy
$[a_i]+[a_j]+[a_k]+[a_l]=[b_n]$.
Solving the resulting simple linear system of equations, it turns out that these can be determined
in terms of the known classes $c_1, -t$  and three arbitrary ones (dubbed here $\chi_{6,7,8}$), which
will be treated as free parameters~\cite{Dudas:2010zb}.
Each matter curve is associated with a defining equation involving products of $a_i$'s
and, as such, it belongs to a specific homological class which subsequently is used to determine
the flux restriction on it. If  ${\cal F}_Y$ represents the hypercharge flux, we will require
 the vanishing of ${\cal F}_Y\cdot c_1=  {\cal F}_Y\cdot(-t)=0$, so
that all  can be expressed in terms of three free (integer parameters) defined by
the restrictions
\be
N_7={\cal F}_Y\cdot \chi_7,\;   N_8={\cal F}_Y\cdot \chi_8,\;   N_9={\cal F}_Y\cdot \chi_9\,.
\ee

\noindent
To construct a specific model, we start by assuming that a suitable $U(1)_X$ flux (where the abelian factor
$U(1)_X$ lies outside $SU(5)$ GUT) generates chirality for the $10$ and $\bar 5$ representations.  Next, the
hypercharge  flux breaks $SU(5)$  down to the SM and, at the same, time it splits the $10, \overline{10}$
and $5,\bar 5$'s  into different numbers of SM multiplets. If some integers  $M_{10}, M_5$ are
 associated with the $U(1)_X$ flux, and some linear combination  $N_y$ of $N_{7,8,9}$ represents the
corresponding   hyperflux piercing  a given  matter curve, the 10-plets and 5-plets  split according to:
\be
    {10}_{t_{i}}=
                            \left\{\begin{array}{ll}{\rm
Representation}&{\rm flux\, units }\\
                         n_{{(3,2)}_{1/6}}-n_{{(\bar 3,2)}_{-1/6}}&=\;M_{10}\\
                        n_{{(\bar 3,1)}_{-2/3}}-n_{{(
3,1)}_{2/3}}&=\;M_{10}-N_y\\
                        n_{(1,1)_{+1}}-n_{(1,1)_{-1}}& =\;M_{10}+N_y\\
                            \end{array}\right.\,,
                            \label{F10}
                            \ee
                            \be
                            {5}_{t_{i}}=
                            \left\{\begin{array}{ll}{\rm
Representation}&{\rm flux\, units }\\
                         n_{(3,1)_{-1/3}}-n_{(\bar{3},1)_{+1/3}}&=\;M_5\\
                        n_{(1,2)_{+1/2}}-n_{(1,2)_{-1/2}}& =\;M_5+N_y\,\cdot\\
                            \end{array}\right.
                            \label{F5}
  \ee
As already discussed, depending on the  restrictions of the flux on the matter curves $\Sigma_j$, there are certain
conditions on the corresponding hypercharge flux,  denoted as $N_{y_j}$ (for the specific matter curve  $\Sigma_j$).
These are deduced from the topological properties of the coefficients $a_i$ as well as the fluxes.

For a given choice of the flux parameters $M_i, N_{y_j}$,   the most general spectrum and its properties under
the assumption of a $Z_2$ monodromy are exhibited in Table~\ref{genspec}. The first column shows the available matter
 curves and the assumed chiral state propagating on it. The chirality is fixed by the specific choice of $M_i, N_{y_j}$
 flux coefficients shown in the last  two columns of Table~\ref{genspec}. The second column shows the  `charge' assignments,
$\pm t_i$ for the 10-plets, and $\pm (t_i+ t_j), \pm (t_i- t_j)$ for 5-plets and singlets respectively.
For this particular arrangement, the structure of the fermion mass matrices  exhibits a hierarchical form, consistent with
 the experimentally measured masses and mixings~\cite{Leontaris:2010zd}. In the present work, we will explore other interesting
 phenomenological implications of this model. The defining equations are shown in  the fourth column where, for
 brevity, the notation $a_{ijk...}=a_ia_ja_k\cdots $ is used. The next column indicates the homologies, the sixth column
  their associated integers expressing the restrictions of flux  on the corresponding matter curves, and the last column
lists  a choice of $M_i$ values consistent with a chiral $SU(5)$ spectrum. Notice that the flux integers are subject to the
 restrictions~\cite{Dudas:2010zb} $N=N_7+N_8+N_9$ and $\sum_iM_{5_i}+\sum_jM_{10_j}=0$.  In the minimal case  $n=0$
  and there are no extra $5+\bar 5$ pairs. Furthermore,  the multiplicities $M_{ij}, M_{\delta}$ of singlet fields are
  not  determined in the context of the spectral cover  and are left arbitrary.

\begin{table}[tbp] \centering%
	\begin{tabular}{|ll|c|c|c|c|c|}
		\hline
	Curve&Field&$U(1)_i$& defining eq. & homology& $U(1)_Y$-flux&$U(1)$-flux\\
		\hline
		$\Sigma_{10^{(1)}}$:&$10_3$&  $t_{1}$&$a_1$ & $\eta-2c_1-{\chi}$&$ -N=0$ &$M_{10_1}=1$\\ %\hline
		$\Sigma_{10^{(2)}}$:&$10_1$& $t_{3}$&$a_4 $ &$-c_1+\chi_7$&$ N_7=-1$ &$M_{10_2}=1$\\% \hline
		$\Sigma_{10^{(3)}}$:&$10_2$& $t_{4}$&$a_6 $ &$-c_1+\chi_8$&$ N_8=1$ &$M_{10_3}=1$\\% \hline
		$\Sigma_{10^{(4)}}$:&$10_2'$& $t_{5}$&$a_8$&$-c_1+\chi_9$&$ N_9=0$ &$M_{10_4}=0$\\ %\hline
		$\Sigma_{5^{(0)}}$:&$5_{H_u}$& $-2t_{1}$&$a_{578}+a_{479}+a_{569} $ & $-c_1+{\chi}$&$ N=0$ &$M_{5_{H_u}}=1$\\ %\hline
		$\Sigma_{5^{(1)}}$:&$\bar 5_2$& $t_{1}+t_3$&$a_1-c(a_{478}+a_{469})$&$\eta -2c_1-{\chi}$&$ -N=0$ &$M_{5_1}=-1$\\% \hline
		$\Sigma_{5^{(2)}}$:&$\bar 5_3$& $t_{1}+t_4$&$a_1-c(a_{568}+a_{469}) $ &$\eta -2c_1-{\chi}$&$ -N=0$ &$M_{5_2}=-1$\\% \hline
		$\Sigma_{5^{(3)}}$:&$5_x$& $-t_{1}-t_5$&$a_1-c(a_{568}+a_{478}) $  &$\eta -2c_1-{\chi}$&$ -N=0$ &$M_{5_3}=n$\\% \hline
		$\Sigma_{5^{(4)}}$:&$\bar 5_1$& $t_{3}+t_4$& $a_{56}+a_{47} $&$-c_1+{\chi}-\chi_9$&$N-N_9=0$ &$M_{5_4}=-1$\\ %\hline
		$\Sigma_{5^{(5)}}$:&$\bar 5_{H_d}$& $t_{3}+t_5$&$a_{58}+a_{49} $ & $-c_1+{\chi}-\chi_8$&$ N-N_8=-1$ &$M_{5_{H_d}}=0$\\% \hline
		$\Sigma_{5^{(6)}}$:&$\bar 5_{\bar x}$& $t_{4}+t_5$&$ a_{78}+a_{49}$ &$-c_1+{\chi}-\chi_7$&$ N-N_7=1$ &$-n-1$\\
		                   &$\theta_{12}$& $0$&$ -$ &$-$&$ -$ &$M_{12}$\\
		$\Sigma_{5^{(6)}}$:&$\,\theta_{ij}$& $t_i-t_j$&$ -$ &$-$&$ -$ &$M_{ij}$\\
		         &$\theta_{\delta}$& $0$&$ -$ &$-$&$ -$ &$M_{\delta}$ \\\hline		
	\end{tabular}%
	\caption{Field content under $SU(5)\times U(1)_{t_i}$, their homology class and flux restrictions.
	 For convenience, only the properties of $10, 5$ are shown. $\ov{10},\ov{5}$ are characterised by
	 $t_i\ra -t_i$. Note that the fluxes satisfy $N=N_7 +  N_8 + N_9$ and $\sum_iM_{10_i}+\sum_jM_{5_j}=0$,
     while  ${\chi}=\chi_7 +  \chi_8 + \chi_9$.}
	\label{genspec}
\end{table}

\section{Spectrum of the effective low energy theory}

A comprehensive classification  of the resulting spectrum  is shown in Table~\ref{modA} where, in the
first column, the $SU(5)$ properties are shown. The third column shows the accommodation of the SM representations
with their corresponding `charges' given in column 2. Column 4 includes the exotics which, for
the specific choice of parameters, involves the triplet pair $D+ D^c$ and, in principle, $n $ copies
of $5+\bar 5$ representations. In the minimal case we set $n=0$, but perturbativity allows values up to $n\le 4$.
In the modified version of the model  we allow for $n\ne 0$ and explore the phenomenological implications.
note that restrictions on the number of vector-like 5-plets  arise when the model is embedded in an $E_6$ 
framework~\cite{Callaghan:2011jj}-\cite{Callaghan:2013kaa}.
 In the last column of the  Table, we have also introduced a $Z_2$ matter parity to the MSSM field as well as the singlets.

Before proceeding with the main part of our paper we present a few remarks about R-parity in supersymmetric models.
A discrete $Z_2$ R-parity is often invoked in four dimensional supersymmetric SU(5) models in order to eliminate
rapid proton decay mediated by the supesrymmetric partners of the SM quarks and leptons. If left unbroken, this discrete
symmetry also yields an attractive candidate for cold dark matter, namely the lightest neutralino. It is perhaps worth
noting that this $Z_2$ symmetry naturally appears if we employ an $SO(10)$ GUT which is broken down to
$ SU(3)_c \times  U(1)_{em}$ by utilizing only tensor representations~\cite{Kibble:1982ae}.

The question naturally arises: how do string theory based unified models avoid rapid proton decay? In the ten-dimensional $ E_8 \times  E_8$
heterotic string framework~\cite{Candelas:1985en}, the compactification process utilizes
Calabi-Yau manifolds which typically yields non-abelian discrete symmetries  that may contain the desired R-parity~(\cite{Lazarides:1989yz} and references therein.)

In F-theory  models  discrete symmetries including R-parity may  arise  from a variety of sources. They can emerge
from  Higgsing  $U(1)$ symmetries in F-theory compactifications,  or from a non-trivial Mordell-Weil group associated with the rational sections
of the elliptic fibration, first invoked in ~\cite{Braun:2014oya} and further discussed in several works
including~\cite{Lin:2015qsa,Anderson:2014yva,Klevers:2014bqa,Cvetic:2015moa}. More generally, $Z_n$ symmetries are associated with Calabi-Yau
 manifolds whose geometries are associated with the Tate-Shafarevich group~\cite{Cvetic:2016ner}.
 Finally, they may appear as geometric properties of the construction in the spectral cover picture~\cite{Karozas:2014aha}.
Based on the existence of such possibilities, in the present model we implement the notion of $R$-parity assuming that it is associated
with some  symmetry of geometric origin.

 \subsection{Matter curves and Fermion masses}

Returning to the description of the emerging effective model, for further clarification we include a few more details.
Initially, in the covering theory there are five matter curves~\footnote{Recall from~(\ref{E8To5}),
$\Sigma_{10_i}, i=1,2,\dots,5$ that the 10-plets transform in the fundamental and 5-plets in the antisymmetric
representation of $SU(5)_{\bot}$.} but due to monodromy $Z_2: t_1=t_2$, two of them are identified and thus they are reduced
 to four. Similarly, the ten $\Sigma_{5_{t_i+t_j}}$ reduce to seven matter curves. Furthermore, there are 24
 singlets from the decomposition of the adjoint of $SU(5)_{\perp}$ denoted with $\theta_{ij}, i,j=1,2\dots, 5$,
  and 20 of them live on matter curves defined by $t_i-t_j$ while four are `chargeless'.
However, because of  the  $Z_2$ monodromy  among the various identifications, $\theta_{i1}\equiv \theta_{i2}$
and $\theta_{1j}\equiv \theta_{2j}$,  the following two singlets:
\be  \theta_{12}=\theta_{21}\to S
 \label{MonSinglet}
\ee
are equivalent to one singlet $S$ with  zero charge. The remaining singlets with non-zero `charges' are
\[ \theta_{13},\, \theta_{14},\, \theta_{15},\, \theta_{34},\, \theta_{35},\, \theta_{45},\, {\rm and}\;
\theta_{31},\, \theta_{41},\, \theta_{51},\, \theta_{43},\, \theta_{53},\, \theta_{54}\]
The following singlets acquire  non-zero vevs which help in realising the desired fermion mass textures:
		\be
		\langle\theta_{14}\rangle\equiv V_1\equiv v_1 M_{GUT} \ne 0, \langle\theta_{15}\rangle\equiv V_2
	\equiv v_2 M_{GUT}	\ne 0, \langle\theta_{43}\rangle\equiv V_3 \equiv v_3 M_{GUT}\ne 0\,.\label{singlevevs}
		\ee 	
All other  singlets  (designated with $\theta^{\perp}_{ij}$ in Table~\ref{modA}) have zero vevs.
Using the  SM Higgs and singlet  vevs given by~(\ref{singlevevs}), we obtain  hierarchical
 quark and charged mass  textures
\be
  M_u\propto \left(
  \begin{array}{ccc}
   v_1^2 v_3^2 & v_1^2 v_3 & v_1 v_3 \\
   v_1^2 v_3 & v_1^2 & v_1 \\
   v_1 v_3 & v_1 & 1 \\
  \end{array}
  \right)\langle H_u\rangle ,\;
  M_{d,\ell}=\left(
  \begin{array}{ccc}
   v_1^2 v_3^2 & v_1 v_3^2 & v_1 v_3 \\
   v_1^2 v_3 & v_1 v_3 & v_1 \\
   v_1 v_3 & v_3 & 1 \\
  \end{array}
  \right)\langle H_d\rangle\,,
  \ee
where, the Yukawa couplings are suppressed for simplicity.

 \subsubsection{Neutrino sector}

  The  tiny  masses accompanied by the relatively  large mixings of the neutrinos, as indicated by various experiments,
   can find a plausible solution
  in the context of the see-saw mechanism and the existence of family symmetries. In the present F-$SU(5)$ GUT  model, the
  SM singlet fields such as $\theta_{ij}$ form Yukawa terms invariant under the additional family  symmetries described above
  and could be  the natural candidates for the right  handed neutrinos.  Furthermore, observing that  the right-handed
  neutrino mass scale is of the order of the Kaluza-Klein scale in string compactifications, a minimal scenario would be to
  associate the right handed neutrinos with the KK-modes~\cite{Bouchard:2009bu} of these singlet fields,
  $\theta_{ij}^{KK}\to N_{R}$. An obstruction to this interpretation  is that in the covering theory these singlets  $\theta_{ij}$
  transform in the complex  representation, so that $\theta_{ij}^{KK}= N_R, \theta_{ji}^{KK}= N^c_R$ and the mass term becomes
  $M_{KK}N_RN_R^c$, but there are no corresponding Dirac mass terms for both $N_R, N_R^c$.
  However, in the quotient theory under the $Z_2$ monodromy $t_1=t_2$, the  KK-modes $\theta_{12}^{KK}\equiv \theta_{21}^{KK}$
  transform in the real representation, so that for any KK-level the corresponding modes $N_{R_k}=N_{R_k}^{c}\to \nu_k^c$
  are identified and a see-saw mechanism is possible.  Hence,  the non-renormalisable term $5_{-t_1-t_2}\bar 5_{t_1+t_4}\theta_{14}\theta_{21}^{KK}$
  under the $Z_2$ monodromy is identified with $5_{-2t_1}\bar 5_{t_1+t_4}\theta_{14}\theta_{21}^{KK}\to 5_{h_u}\bar 5_3\theta_{14}\nu^c$ and so on.
  Therefore, under the above assumptions, the KK-modes corresponding to right-handed neutrinos couple to the following combination
  of the left-handed neutrino components
  \be
   5_{H_u}(\bar 5_1 \theta_{14}^2 \theta_{43}+\bar 5_2\theta_{14}\theta_{43}+\bar 5_3\theta_{14})\,.\label{MajNKK}
   \ee
 The interesting fact is that  the right-handed neutrinos are associated with a specific class of  wavefuctions~\cite{Bouchard:2009bu} 
 such that the emerging mass hierarchy is milder than that of the charged leptons and quarks. It is shown that the  mass matrix obtained 
 this way~\cite{Bouchard:2009bu} can accommodate the two large mixing angles observed in atmospheric and solar neutrino experiments.
     \begin{table}[tbp] \centering%
  	\begin{tabular}{|l|c|c|c|c|}
  		\hline
  		Irrep&$U(1)_i$&SM spectrum & Exotics&$R$-parity\\
  		\hline
  		$\,10_1$& $t_{3}$&$Q_1, u_1^c, u_2^c  $ &$-$&$-$\\% \hline
  		$\,10_2$& $t_{4}$&$Q_2, e^c_1, e^c_2 $ &$-$&$-$\\% \hline
  		$\,10_3$&  $t_{1}$&$Q_3, u_3^c, e_3^c$ & $-$&$-$\\ %\hline
  		$\,\bar 5_1$& $t_{3}+t_4$& $d^c_1, \ell_1 $ &$-$&$-$\\ %\hline
  		$\,\bar 5_2$& $t_{1}+t_3$&$d^c_2, \ell_2$&$-$&$-$\\% \hline
  		$\,\bar 5_3$& $t_{1}+t_4$&$d^c_3, \ell_3$ &$-$&$-$\\ \hline
  		$\,5_{H_u}$& $-2t_{1}$&$H_u $ & $D$&$+$\\ %\hline
  		$\,\bar 5_{H_d}$& $t_{3}+t_5$&$H_d $ & $-$&$+$\\% \hline
  		$\,5_x$& $-(t_{1}+t_5)$&$-$  &$ (H_{u_i}, D_i)_{i=1,...,n}$&$+$ \\ % \hline
  		$\,\bar5_{\bar x}$& $t_{4}+t_5$&$-$ &$ D^c+  (H_{d_i}, D_i^c)_{i=1,...,n}$&$+$\\
  		 \hline
  		$\theta_{12,21}$& $0$&$ $ &$  S \;({\rm singlet})$&$-$\\
  		$\,\theta_{14}$& $t_1-t_4$&$ {}$ &$\langle\theta_{14}\rangle =V_1\equiv v_1 M_{\tiny GUT} $&$+$\\
  		$\,\theta_{15}$& $t_1-t_5$&$ {}$ &$\langle\theta_{15}\rangle =V_2\equiv v_2 M_{GUT}$&$+$\\
  		$\,\theta_{43}$& $t_4-t_3$&$ {}$ &$ \langle\theta_{43}\rangle=V_3\equiv v_3 M_{GUT} $&$+$\\
  		$\theta^{\perp}_{ij}$&$t_i-t_j$&${}$ & $\langle\theta^{\perp}_{ij}\rangle=0$&$+$\\
  	%	$\theta_{\delta}$& $0$&$ {}$ &moduli& \\	
  		\hline		
  	\end{tabular}%
  	\caption{Field  content under $SU(5)\times U(1)_{t_i}$. The third column shows the MSSM spectrum and the fourth column displays
  	the predicted	exotics. 	The R-parity assignments appear in the last column. We use assignments	${10}_{t_i}, {5}_{-t_i-t_j}$
  	 whith	$\ov{10},\ov{5}$ characterized by opposite values, $t_i\ra -t_i$ etc.  The fluxes eliminated components of the $SU(5)$
  	  multiplets, giving rise	to incomplete representations. There are also $n$ copies, of	$5+\bar 5$ multiplets. }
  	\label{modA}
  \end{table}
  \begin{figure}[!bth]
    \centering
    \includegraphics[scale=0.45]{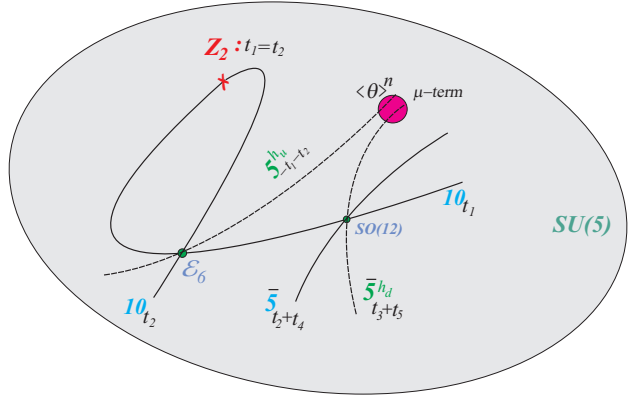}
    \label{TBY}
    \caption{The trilinear  top and bottom  Yukawa couplings at the triple intersections of the
    matter curves with symmetry enhancements  ${\cal E}_6$ and $SO(12)$  respectively.
     Under a $Z_2$ monodromy we obtain identifications such as $10_{t_1}=10_{t_2}$
    so that a `diagonal' top Yukawa coupling can be realised. A $\mu$-term emerges only
    from non-renormalisable (suppressed) contributions. $\langle\theta\rangle^n$ stands for
    the ratio of singlet vevs divided by the high (compactification) scale.}
    \end{figure}

\subsection{Mass terms for the doublets and triplets}

Returning to the content of Table~\ref{modA}, we observe that there is still freedom to
accommodate additional vector-like 5-plets which respect all the required conditions.
Hence, aiming to accommodate potential diphoton resonances and other possible experimental signatures
of exotic matter beyond the MSSM spectrum, in the present construction
 we assume the existence of $5+\bar 5$ pairs and discuss  possible   implications of the exotic
states. As already explained, the $Z_2$ monodromy allows a tree-level coupling for the
top quark $10_310_3 5_{H_u}$. Furthermore, from the specific accommodation of the fermion
generations listed in Table 2, we observe that a tree-level coupling for the bottom
quark is also available. A geometric perspective of the Yukawa couplings in the internal manifold
is depicted in figure~\ref{TBY}. All other mass entries are  generated from non-renormalisable terms~\cite{Leontaris:2010zd}.

\noindent
Regarding  the 5-plets accommodating the MSSM Higgs,  we observe that the  flux splits the doublet
from the triplet in the Higgs sector.  As a result, the  MSSM $\mu $ term
  \be
  \frac{\theta_{14} \theta_{43}\theta_{15}}{M_{GUT}^2} \, \bar 5_{t_3+t_5}5_{-2 t_1}
   \to  \frac{V_1V_2V_3}{M_{GUT}^2}H_uH_d\to \mu H_uH_d. \label{Muterm}
   \ee
   does not involve masses for the triplet fields. Fermion mass hierarchies require at least   that the singlet vev
    $V_1=\langle \theta_{14}\rangle\gtrsim {\cal O}(10^{-1})M_{GUT}$,  so that
    the MSSM $\mu$ parameter can be kept light for $v_2\cdot v_3\ll v_1$.

    In the general case, we need to take into account the extra doublet pairs emerging from the 5-plets remaining in the
    zero-mode spectrum.   As an illustrative example, we take only one additional vector-like pair of 5-plets, that is $n=1$.
    In this case  the available coupling are
    \[ 5_{H_u} \bar 5_{H_d}	\theta _{14}\theta _{43} \theta _{15}/M_{GUT}^2 +  5_{H_u} \bar 5_{\bar x}	\theta _{14}\theta _{15}/M_{GUT}
    + 5_{x} \bar 5_{H_d}	\theta _{14}\theta _{43}/M_{GUT}+ 5_{x} \bar 5_{\bar x}	\theta _{14}\,.
    \]
 The Higgs mass matrix in the basis ${\cal L}\supset {\tiny (H_d,H_d') M_H \left(\begin{array}{c}H_u\\ H_u'\end{array}\right)}$ is
\ba
M_H&\propto &	V_1 \left(
\begin{array}{ll}
	v _{3} v _{2} &  v _{3}  \\
     v_{2} &1  \\
\end{array}
\right)\,,
\label{Mh2}
\ea
where the Yukawa couplings are suppressed to avoid clutter.  This implies a light Higgs mass term
$\mu\sim V_1v_2 v_3$ and a heavy one $M_H\sim V_1$.

 \noindent
      The triplet mass terms emerge from different couplings
   \be
 \theta_{14}\theta_{15} \, \bar 5_{t_4+t_5}5_{-2 t_1}/M_{GUT} + \epsilon\theta_{14} \,
 \bar 5_{t_4+t_5}5_{- t_1-t_5}\to
 \theta_{14}\theta_{15} \, \bar 5_{\bar x}5_{H_u}/M_{GUT} + \epsilon\theta_{14} \,
 \bar 5_{\bar x}5_{x}\,\cdot
    \ee
  Hence, written in a matrix form  $${\cal L}_D \supset \tiny{ \left(	5_{H_u},	5_{x} \right)
  	M_D
  	\left(
  	\begin{array}{c}
  \bar 5_{H_d} \\
  \bar 5_{\bar x} \\
  	\end{array}
  	\right)},$$
 where 	the triplet mass matrix is
$ M_D={\tiny V_1\left(
 \begin{array}{ll}
 	 v _{2} &	\eps   \\
 \eps' v _{2} & 1  \\
 \end{array}
 \right)}$,
and the parameters $\eps\simeq \eps'\lesssim 1$ stand for corrections when more than one matter multiplets are on the same matter curve.
 The  eigenmasses also  depend on the singlet vevs and will be discussed in conjunction with proton decay in the subsequent sections.

 In addition to these  superpotential couplings, the vector pairs $5+\bar 5$ generate superpotential terms with the matter fields
  \be 10_3 \bar 5_{\bar x}\bar 5_2, 10_3 \bar 5_{\bar x}(\bar 5_1\theta_{14}+\bar 5_3\theta_{34}),
  10_1 \bar 5_{\bar x}(\bar 5_1\theta_{14}\theta_{43}+\bar 5_2\theta_{43}+\bar 5_3)\theta_{14},
  10_2 \bar 5_{\bar x}(\bar 5_1\theta_{14}+\bar 5_2+\bar 5_3\theta_{34})\theta_{14}\,\cdot
  \label{RPV}
  \ee
where the non-renormalisable terms are assumed to be scaled by appropriate powers of $M_{GUT}$.
 In the next sections we will explore possible phenomenological consequences of~(\ref{RPV}).
However, we note that it is feasible to eliminate  such couplings from the lagrangian
by  introducing  a different  R-parity assignment for the colour triplets.

 \section{ Gauge Coupling Unification}

   The presence of additional vector-like pairs of colour triplets and higgsinos with masses in the TeV range  affect the
    renormalisation     group running  of the gauge couplings and the fermion masses.  The existence of  complete   $5+\bar 5$ 
     $SU(5)$ multiplets  at the TeV scale may enhance processes that could be observed in future searches, while they can be consistent with perturbative  gauge  coupling unification as long as their number is less than four. Threshold corrections from Kaluza-Klein (KK) modes and fluxes play a significant r\^ole~\cite{Donagi:2008kj} too.
     Under certain circumstances~\cite{Leontaris:2011tw}, (for example when the matter fields are localised on genus one surfaces)
    the KK threshold effects can be universal, resulting to a   common shift of the gauge coupling constant at the GUT scale.
    This has been analysed in some  detail in ref~\cite{Leontaris:2011tw} and will not  be elaborated further. 
   However, in F-theory constructions, there are additional corrections associated with  non-trivial line bundles~\cite{Blumenhagen:2008aw,Leontaris:2009wi}). More precisely, assuming that the   $SU(5)$ is generated by D7-branes wrapping  a del-Pezzo surface,
    gauge flux quantization condition~\cite{Blumenhagen:2008zz}   implies that D7-branes are associated with a non-trivial line
    bundle ${\cal L}_a$.   On the other hand,  the breaking of $SU(5)$ occurs with a non-trivial  hypercharge flux ${\cal L}_Y$
     supported on the del Pezzo surface, (but with a trivial restriction on the Calabi-Yau fourfold so that the associated  gauge boson  remains massless). The flux threshold corrections to the gauge couplings  associated with these two line bundles can be computed by  dimensionally reducing the Chern-Simons action.  If we define
    \be
     y=\frac 12 {\rm Re}S \int c_1^2({\cal L}_a),\;  x  =-\frac 12 {\rm Re}S \int c_1^2({\cal L}_Y)~,
    \ee
     where, $c_1({\cal L})$ denotes the first Chern class of the corresponding line bundle and
      $S=e^{-\phi}+iC_0$ is the axion-dilaton field (and $g_{_{IIB}}=e^{\phi}$),
      the flux corrections to the gauge couplings are  expressed as follows
     \ba
     \frac{1}{a_3(M_U)}&=&\frac{1}{a_U}-y\label{A3}\\
      \frac{1}{a_2(M_U)}&=&\frac{1}{a_U}-y+x\label{A2}\\
       \frac{1}{a_1(M_U)}&=&\frac{1}{a_U}-y+\frac 35x\label{A1}\,,
   \ea
    where $a_U$ represents  the unified gauge coupling.
    From ~(\ref{A3}-\ref{A1}) we observe that the corrections from the ${\cal L}_a$ line bundle are
    universal and therefore $y$ can be absorbed in a redefinition of $a_U$. On the other hand,
   hypercharge flux thresholds expressed in terms of $x$, are not universal and destroy  the
   gauge coupling unification at the GUT scale  $M_U$.  Notice that in order to eliminate the
    exotic bulk states $(3,2)_5+(\bar 5,2)_{-5}$ emerging from the decomposition of ${\bf 24}$,
    we need to impose $ \int c_1^2({\cal L}_Y)=-2$, and therefore we find the simple form
    $x= e^{-\phi}=\frac{1}{g_{_{IIB}}}$. The value of the gauge coupling  splitting has important
    implications on the mass scale of the color triplets discussed in the previous section. In the
    following we will explore this relation within the matter and Higgs field context of the present model.

   We  assume that the color triplets  $D+ D^c\in 5_H+\bar 5_H$ receive masses at a scale $M_X$, while
   the complete $5+\bar 5$ extra multiplets obtain masses at a few TeV. The renormalisation group equations
   take the form
   \be
   \frac{1}{a_i(M_U)}=\frac{1}{a_i(M_U)}+\frac{b_i^x}{2\pi}\log\frac{M_U}{M_X}+ \frac{b_i}{2\pi}\log\frac{M_X}{\mu}~\cdot
   \ee
    It can be readily checked that the GUT values of the gauge coupling satisfy
     \be
     \frac 53 \frac{1}{a_1(M_U)}= \frac{1}{a_2(M_U)}+\frac 23 \frac{1}{a_2(M_U)}~\cdot\label{GUTrel}
    \ee
   Assuming $n_D$ pairs  of $ (D+D^c) $ and   $n_V$ vector-like 5-plets,  the beta functions are
    \ba
    b_3^x=-3+n_V+n_D,\; b_2^x=1+n_V,\; b_1^x=\frac{33}{5}+\frac 25n_D+n_V\\
     b_3=-3+n_V,\; b_2=1+n_V,\; b_1=\frac{33}{5}+n_V~\cdot
    \ea
   Using (\ref{A3},\ref{A2}) and (\ref{GUTrel}) we find
   \be
    \log\frac{M_U}{M_X}= \frac{2\pi}{\beta_x}\frac{1}{\cal A}-\frac{\beta}{\beta_x} \log\frac{M_X}{\mu}\label{MUscale}
    \ee
     where we introduced the definitions
     \ba
     \beta&=& \frac{5}{3}(b_1-b_3)+(b_3-b_2)\\
      \beta_x&=& \frac{5}{3}(b^x_1-b^x_3)+(b^x_3-b^x_2)\\
     \frac{1}{\cal A}&=& \frac{5}{3}\frac{1}{a_1}-\frac{1}{a_2}-\frac 23 \frac{1}{a_3}=\frac{1-2\sin^2\theta_W}{a_e}-\frac 23 \frac{1}{a_3}~\cdot
     \ea
    Notice  that for the particular spectrum,  $\beta_x,\beta$ are equal, $\beta_x=\beta=12$,
     and independent of the number of multiplets $n_D$ and $n_V$.
     Then, from (\ref{MUscale}) we find that the unification  scale is
   \be
   M_U=e^{\frac{2\pi}{12{\cal A}}}M_Z\approx 2.04\times 10^{16} GeV\,, \label{MUnum}
   \ee
   i.e., independent of $n_V, n_D$ and the intermediate scale $M_X$.

   To unravel the relation between the scale $M_X$ and the  parameter $x$, we proceed as follows. First, we subtract (\ref{A2})
   from (\ref{A3})
   \ba
   x&=& \frac{1}{a_2}-\frac{1}{a_3}+\frac{b_3^x-b_2^x}{2\pi}\frac{M_U}{M_X}+\frac{b_3-b_2}{2\pi}\frac{M_X}{\mu}\label{a23x}\\
   &=& \frac{1}{a_2}-\frac{1}{a_3}-\frac{4-n_D}{2\pi}\frac{M_U}{M_X}-\frac{4}{2\pi}\frac{M_X}{\mu}~\cdot\nn
   \ea
   Using (\ref{MUscale}) and the fact that in our model $n_D=1$, we find
   \be
   \log\frac{M_X}{\mu}= {2\pi}\left(\frac{6\sin^2\theta_W-1}{4a_e}-\frac 56\frac{1}{a_3}-x\right)~\cdot\label{MXscale}
   \ee
   This determines the relation between the  parameter $x=e^{-\phi}$ and the scale $M_X$ where the Higgs
   triplets become massive.  We can use the expression for $M_U$ to express the $M_X$ scale as follows
   \be
   \log\frac{M_X}{M_U}=
   {2\pi}\left(\frac{5\sin^2\theta_W-1}{3a_e}-\frac 79\frac{1}{a_3}-x\right)~\cdot\label{MXUscale}
   \ee

\begin{figure}[!bth]
      \centering
      \includegraphics[scale=0.9]{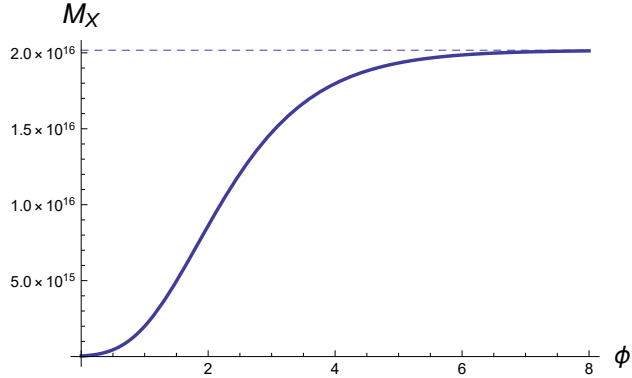}
      %\;\;
      %\includegraphics[scale=0.45]{LogMXU}
      \caption{Variation of $M_X$  scale with respect to the dilaton field.
      For the chosen range of $\phi\in (0,\infty)$, (strong $g_{_{IIB}}$ coupling regime)
      there is a lower bound $M_X\sim 10^{13}$ GeV.  }
           \label{MXdilaton0}
      \end{figure}

\begin{figure}[!bth]
       \centering
       \includegraphics[scale=0.83]{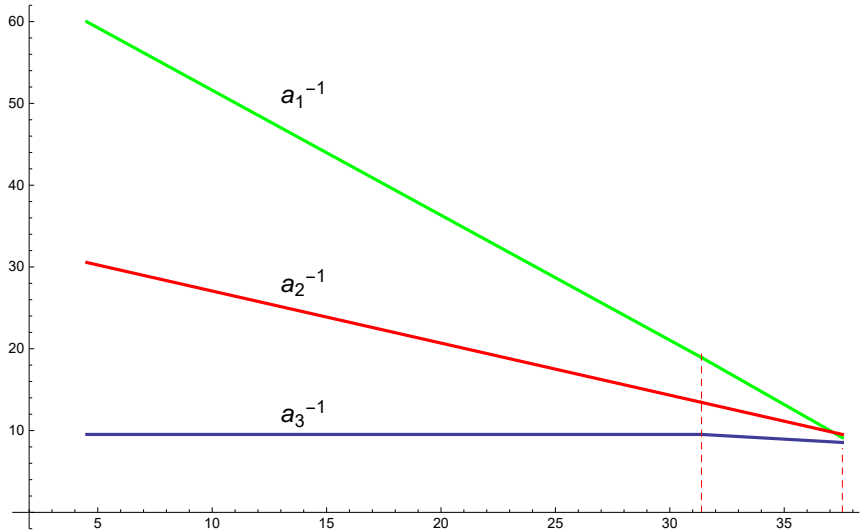}
       \caption{Gauge coupling running  in the presence of flux thresholds and the triplet's decoupling scale $M_X$. }
        \label{RGEplot}
       \end{figure}

   To determine the value of  the GUT coupling $a_U$ we use (\ref{A3},\ref{MUscale}) and (\ref{a23x}) to find
   \be
   \frac{1}{a_U}+x= \frac{1}{a_2}-\frac{b^x_2}{\beta_x}\frac{1}{\cal A}= \frac{1}{a_2}-\frac{1+n_V}{12}\frac{1}{\cal A}~\cdot\label{aUa}
   \ee
   For the present application, we allow three pairs of 5-plets, $n_V=3$, and we obtain the relation
   \ba
   \frac{1}{a_U} %&=& \frac{1}{a_2}-\frac{1}{3}\frac{1}{\cal A}-x\nn\\
   &=&\frac{5\sin^2\theta_W-1}{3 a_3}+\frac{2}{9}\frac{1}{a_3}-x~\cdot\label{aUb}
   \ea
   Substitution of (\ref{aUb}) in (\ref{MXUscale}) gives an elegant and very suggestive formula:
   \ba
   M_X&=& e^{2\pi\left(\frac{1}{a_U}-\frac{1}{a_3}\right)} M_U~\cdot
   \ea
   We observe that in order to have $M_X\le M_U$, we always need $a_U\ge a_3\approx \frac{1}{8.5}$.
   We depict the main results in the figures that follow. In fig.\ref{MXdilaton0} we show the variation
   of the color triplets' decoupling scale versus  the range of values of the dilaton and, in fig.\ref{RGEplot},
   we plot the inverse SM gauge couplings taking into account the thresholds of the color-triplets.

   \subsection{RGEs for Yukawa Couplings}

   These  modifications in the gauge sector and,  in particular, the large $g_U$ value   compared to that of the standard MSSM unification scenario  ($g_U\sim 1/25$ in MSSM) are expected
  to have a significant impact on   the evolution  of the Yukawa couplings.  On the other hand, in  F-theory constructions
  the Yukawa coupling strengths at the unification scale are computed analytically and can be expressed in terms of the geometric properties
  of the internal six-dimensional compact space and the fluxes of the particular construction.   For the sake of argument, we
  assume  that
  all three $5+\bar 5$    surplus matter fields receive masses in the TeV range,  with $\tan\beta$ values $\sim 48-50$ and  $M_{GUT}\sim
  2\times 10^{16}$ GeV. Then, according to~\cite{Hebbar:2016gab}, the top mass, in particular, is achieved  for Yukawa coupling  $h_t(M_{GUT})
  \gtrsim 0.35$ which is significantly lower than the value $\sim 0.6$ obtained in the  case of RG running with the beta-functions for the MSSM spectrum.

  \noindent
  Turning now to F-theory predictions, as we have seen, the Yukawa couplings are realised at the intersections of three matter curves.
  The properties of the corresponding matter fields in a given representation $R$ are  captured by the wavefunction $\Psi_R$
  whose profile is  obtained by solving the Equations of Motion (EoM)~\cite{Beasley:2008dc}.
  It is found that the solution exhibits a gaussian profile picked along the matter curve supporting the particular state,
  $ \Psi_R \propto f(  z_i) e^{M_{ij} z_i\bar z_j} $. Here $z_{1,2}$ are local complex coordinates, the `matrix' $M_{ij}$
  takes into account background fluxes,  and $f(z_j)$ is a holomorphic function.
  The value of the Yukawa coupling results from integrating over  the overlapping  wavefunctions.
  Thus, for the up/down Yukawa couplings,
  \be
  h_t \propto  \int \Psi_{10}\Psi_{10} \Psi_{ 5_{H_u}} dz_1\wedge d\bar z_1\wedge dz_2\wedge d\bar z_2,\; h_b \propto  \int \Psi_{10}\Psi_{\bar 5} \Psi_{\bar 5_{H_d}} dz_1\wedge d\bar z_1\wedge dz_2\wedge d\bar z_2\,.
  \label{Ytbtau}
  \ee
  The top Yukawa coupling is realised at the intersection where the symmetry is enhanced to ${ E}_6$, while
  the bottom and $\tau$ Yukawa couplings are associated with triple intersections of $SO(12)$ enhancements.
  We note in passing that the corresponding solution of the EoM
  providing the wavefunction for the up-type quark coupling is rather involved because of the
  monodromy  and must be solved in a non-trivial  background where the notion of  T-brane is required~\cite{Cecotti:2010bp}.
  Using  appropriate background fluxes, we can break ${\cal E}_6$ to
  $SU(5)$,  while the latter can break down to the SM gauge group with the hypercharge flux.
  To estimate the top Yukawa coupling, one has to perform the corresponding integration~(\ref{Ytbtau}). Varying the various flux parameters
  involved in the corresponding wavefunctions, it is found that the  top quark Yukawa takes values in the interval $h_t\sim [0.3-0.5] $,
  in agreement with previous computations~\cite{Font:2013ida,Font:2015slq}, and hence the desired value $h_t\sim 0.35$ can be accommodated.

  In the present approach, the bottom and $\tau$ Yukawa couplings are formed at a different point of the compact space
  where the symmetry enhancement is $SO(12)$. Proceeding in analogy with the top Yukawa, one can adjust
  the flux breaking mechanism to  achieve~\cite{Font:2013ida,Font:2015slq}) the successive breaking to $SU(5)$ and $SU(3)\times SU(2)\times U(1)$.
  Further, for certain regions of the parameter space, one can obtain $h_{b, \tau}$ values in agreement with those
  predicted by the renormalisation group evolution~\cite{CrispimRomao:2016tww}.

 \section{Decay of Vectorlike Triplets}

While analysing the spectrum in  section 4, we have seen that the existence of  vector-like
triplets is a frequently occurring phenomenon. They can be produced in pairs at LHC through
their gauge  couplings to gluons.  However, such exotic particles are not yet observed  and  must decay
through higher dimensional operators through mixing with the MSSM particles.

\noindent
We start with the minimal model by setting $n=0$, in which case the only states beyond the MSSM spectrum are $D^c, D$ found in
the $\bar 5_{\bar x}$ and $5_{H_u}$ respectively.  We will consider the case of their mixing with the third family which
 enhances their  decays, due to the large Yukawa coupling compared to the  two lighter generations.
 Τhe available Yukawa couplings which mix the down-type triplets are
\ba
{\cal W}&\supset &\lambda 10_{t_1} \bar 5_{t_1+t_4} \bar 5_{t_3+t_5}+\lambda_1 10_{t_1} \bar 5_{t_4+t_5} \bar  5_{t_3+t_5}\theta_{15}/M_{GUT}
+\lambda_2 \bar 5_{t_4+t_5}  5_{-2t_1}\theta_{14}\theta_{15}/M_{GUT}\nn\\
&\to &
\lambda 10_{3} \bar 5_{3} \bar 5_{H_d}+\lambda_1 10_{3} \bar 5_{H_d} \bar 5_{\bar x}\, v_2
+\lambda_2 \bar 5_{\bar x}  5_{H_u}\,v_1\,v_2\,,
\ea
where the non-renormalisable terms are scaled by the appropriate powers of the compactification scale or the GUT scale.
These terms generate a mixing matrix of the  third generation down quark and  $D^c,D$ which can be cast in the form
\[{\cal L}_Y\supset \tiny{\left(Q_3, D\right)
M_D
\left(
\begin{array}{c}
b^c \\
D^c  \\
\end{array}
\right)},\; M_D\propto\left(
\begin{array}{cc}
 \frac{\lambda  }{\sqrt{2}}v_d & \frac{ \lambda _1}{\sqrt{2}} v_2 v_d\\
 0 & {\lambda _2 v_1 v_2 } \\
\end{array}
\right)\,, \]
where $v_d$ stands for the down Higgs vev scaled by the GUT scale.
This non-symmetric  matrix $M_D$ is diagonalised by utilizing the left and right unitary matrices
\[M_D^{\delta}= V_L^{\dagger} M_D V_R, \]
implying $${M_D^{\delta}}^2= V_L^{\dagger} M_D M_D^{\dagger}V_L= V_R^{\dagger} M_D^{\dagger}M_D V_R$$
  where
\be
M_DM_D^{\dagger}=
\left(
\begin{array}{cc}
 \frac{1}{2} \lambda ^2 v_d^2+\frac{1}{2}\lambda _1^2 v_2^2  v_d^2 & \frac{
   \lambda _1 \lambda _2}{\sqrt{2}}v_1 v_2^2v_d \\
 \frac{ \lambda _1 \lambda _2}{\sqrt{2}} v_1 v_2^2 v_d&  \lambda _2^2v_1^2v_2^2 \\
\end{array}
\right)
\ee
and
\be
M_D^{\dagger}M_D=
\left(
\begin{array}{cc}
 \frac{1}{2} \lambda ^2 v_d^2 & \frac{1}{2} \lambda \lambda _1  v_2 v_d^2\\
 \frac{1}{2} \lambda \lambda _1  v_2 v_d^2 & \frac{1}{2}\lambda _1^2 v_2^2 v_d^2
  +\lambda _2^2 v_1^2   v_2^2 \\
\end{array}
\right)\,\cdot
\ee
Following standard diagonalisation procedures, in the limit $v_1\gg v_2\gg v_d$, we find
that the left mixing angle is
\be
\tan 2\theta_L =
\frac{\sqrt{2} \lambda _1 \lambda _2 v_1 v_2^2 v_d}{-\frac{1}{2} \lambda ^2
   v_d^2-\frac{1}{2} \lambda _1^2 v_2^2 v_d^2+\lambda _2^2 v_1^2 v_2^2}
     \approx
  \frac{\sqrt{2} \lambda _1 v_d}{\lambda _2 v_1},
   \ee
and for the right-handed mixing   we obtain
\be
\tan 2\theta_R =
 \frac{\lambda  \lambda _1 v_2 v_d^2}{-\frac{1}{2} \lambda ^2 v_d^2+\frac{1}{2} \lambda _1^2
    v_2^2 v_d^2+\lambda _2^2 v_1^2 v_2^2}\approx
    \frac{\lambda  \lambda _1 v_d^2}{\lambda _2^2 v_1^2 v_2}\,.
    \ee
 From these, we find
\[
\tan(2\theta_R)\approx \frac{\lambda  v_d}{\sqrt{2} \lambda _2 v_1 v_2}\tan(2\theta_L)\,\cdot
 \]
 For the assumed hierarchy of vevs we see that the left-mixing prevails.
 The mixing is restricted by CKM constraints and the contributions of the heavy
 triplets to the oblique parameters $S,T$ which have been measured with precision
 in LEP experiments  (For  detailed computations see~\cite{Aguilar-Saavedra:2013qpa}.).
  A rough estimate would give the upper bounds
 $\tan 2\theta_L\sim 0.1, \tan 2\theta_R\sim 0.3 $  which can be easily satisfied for
the  $v_1, v_2$ values  used in this work.

\subsection{Proton decay}

 In this model the dimension-five proton decay R-parity violating tree-level couplings of the form
 $10_f\bar 5_f\bar 5_f$ are absent due to the $t_i$ charge assignments of matter fields. However,
  non-renormalisable terms that could lead to  suppressed baryon and lepton number violating processes may still appear.
 A class  of these operators  have the general structure
  \be \lambda_{eff}  10_i \bar 5_{t_j+t_k}\bar 5_{t_l+t_m},; \lambda_{eff} \sim \langle \theta_{pq}^n\rangle ,
  \;   i,i,j,l,m\ne 5\,,\label{pdeff}
  \ee
where $ \theta_{pq}^n$ represents   products of {singlet fields} required to cancel the  non-vanishing combinations of $t_{i,j...}$ charges.
Notice, however, that for the particular family assignment in this model none of $t_{i,j,k,l,m}$ in~(\ref{pdeff}) is $t_5$ and therefore,
to fulfil the condition $\sum_{k=1}^5t_k=0$ some singlet $\theta_{5s}\equiv 1_{t_5-t_s}$, with $s=1,2,3,4,$ always must be  involved.~\footnote{Notice
however, that all possible  higher order R-parity violating  terms
\[  10_1 \bar 5_1 ( \bar 5_1\theta_{14}\theta_{53}  +\bar 5_2\theta_{53}+ \bar 5_3\theta_{54}) \theta_{13}
+10_1 \bar 5_2 ( \bar 5_2\theta_{43}  +\bar 5_3) \theta_{53} +10_2\bar 5_3\bar 5_3\theta_{54}
\]
can be eliminated due to  the R-parity assignment of the singlets $\theta_{ij}$ shown in Table~\ref{modA}.}
In the present model no singlet of this kind acquires a non-zero vev, namely $\langle \theta_{5s}\rangle \equiv 0$,
and hence dimension four operators are suppressed.

However, as already pointed out, additional Yukawa terms  give rise to new tree-level graphs mediated by color triplets.
Such graphs induce dimension-5 operators of the form $\frac{1}{M_{{eff}}}QQQ\ell$, $\frac{1}{M_{{eff}}}u^cu^cd^ce^c$,
where $M_{{eff}}$ is an effective colour triplet mass
$M_{eff}\ge M_{GUT}\sim 2.0\times 10^{16}$ GeV\cite{Goto:1998qg,Rehman:2009yj,Dermisek:2000hr,Hisano:2013exa}.
Here, because of the missing triplet mechanism described in the previous section,
the $D, D^c$ triplets  develop masses through mixing with other heavy triplets $D_i, D_i'$ emerging from
the decomposition of the additional $5+\bar 5$-pairs.
Besides, several  couplings are realised as higher order non-renormalisable terms  so that,
in practice, an effective triplet mass ${M_{eff}}$ is involved which, with suitable
conditions on the triplet mixing, could be of the order of  the GUT scale.
For the case of the Higgsino exchang diagram, for example, with a Higgsino mass
identified with the supersymmetry breaking scale $M_S$, the proton lifetime is estimated
to be~\cite{Hisano:2013exa}
\be  \tau_p \approx  10^{35} (\sqrt{2}\sin 2\beta)^4 \left(\frac{0.1}{C_R}\right)^2
\left(\frac{M_S}{10^2 TeV}\right)^2\left(\frac{M_{D_{eff}}}{10^{16} GeV}\right)^2\,,
\label{pdlife}
\ee
where the coefficient  $C_R\ge 0.1$,  taking into account the renormalisation group effects on the masses.
From~(\ref{pdlife}) we infer that with an effective triplet mass $\gtrsim M_{GUT}$ and a relatively
high supersymmetry breaking scale, proton decay can be sufficiently suppressed in accordance  with
the Super-Kamiokande bound on the proton lifetime.

To estimate the effects of these operators in this model, we consider the triplet mass matrix
derived in the previous section
 \ba
 M_T&=&\left(
 \begin{array}{ll}
 \lambda	\theta _{14} \theta _{15} &\eps'  \theta _{14} \theta _{15}  \\
 	\eps\theta _{14} & \theta _{14}  \\
 \end{array}
 \right)\,\theta_{14}\; \to\; \left(
  \begin{array}{ll}
  \lambda	v_2 &  \epsilon'\,v_2   \\
  	\eps&1  \\
  \end{array}
  \right)\langle\theta_{14}\rangle, \;
 \label{MD2}
 \ea
with $v_2 =\frac{\langle\theta_{15}\rangle}{M_{GUT}}$ and  $v_1 =\frac{\langle\theta_{14}\rangle}{M_{GUT}}$
as defined in~(\ref{singlevevs}).
As before, the left and right unitary matrices $V_L,V_R$, as well as the eigenmasses are determined by
${M_T^{\delta}}^2= V_L^{\dagger} M_T M_T^{\dagger}V_L= V_R^{\dagger} M_T^{\dagger}M_T V_R$
where, in general,  $M_T M_T^{\dagger}$ and  $M_T^{\dagger}M_T$ are Hermitian but, for simplicity,
we will take to be symmetric, {\small $M^2\sim \left(
 \begin{array}{ll}
 	a & b \\
 b &d \\
 \end{array}
 \right)\langle\theta_{14}\rangle^2$},  with real entries
and triplet eigenmasses {\small $ M^2_{1,2}=\frac 12\left(a+d\pm\sqrt{4 b^2+(a-d)^2}\right)\langle\theta_{14}\rangle^2$}.

\begin{figure}[!bth]
  \centering
 \includegraphics[scale=0.6]{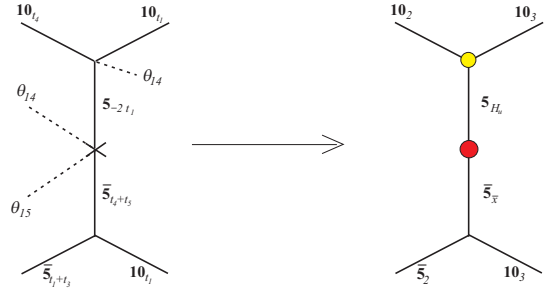}
  \caption{Diagram leading to proton decay. Red and yellow circles   represent non-renormalisable couplings discussed in
  the text.}
\label{PD5}
  \end{figure}

In figure~\ref{PD5} a representative graph is shown  mediated by the colour triplets
leading to the dominant  proton decay mode $p\to K^+\bar \nu$.  The mass insertion (red bullet)  in the graph is
\[  \lambda  \frac{	 \langle\theta _{14} \theta _{15}\rangle}{M_{GUT}}\;\equiv\; \langle\Phi\rangle\,.\]

After  summing over the eigenstates, one finds that the effective mass
involved is
\ba
\frac{1}{M^0_{eff}}\propto  %&=& \sum_j V_{1j}\frac{1}{\fsl p-M_j}\langle\Phi\rangle\frac{1}{\fsl p-M_j}V_{j2}^{\dagger} \nn\\
  % &\to&
    \sum_j V_{1j}\frac{\langle\Phi\rangle}{M_j^2}V_{j2}^{\dagger} \to
 %  &=&
    \left(\lambda \frac{v_2}{v_1}\frac{1}{M_{GUT}}\right)\,\frac{b}{ad-b^2}\,,
\ea
while there is an additional suppression factor $v_1=\langle\theta_{14}\rangle/M$ from the non-renormalisable term (yellow bullet in the graph).
Finally $\frac{1}{M_{eff}}\sim \frac{v_1}{M^0_{eff}}$.

For the $V_L$ mixing, assuming reasonable values for the parameters $\epsilon, \epsilon'<1$, while taking
%\[\frac{1}{ M_{eff}}= \frac{(\epsilon  \lambda +\epsilon' ) \left(1+\epsilon ^2+v_2^2 \left({\epsilon'}^2+\lambda ^2\right)\right)}{v_2^3 (\lambda
%   -\epsilon  {\epsilon'} )^4}\, \frac{v_1}{M_{GUT}}\]
$v_1\sim O(10^{-1})$ and $\lambda\sim 1$ we find
\[ M_{eff} \sim \frac{v_2}{v_1} M_{GUT}\,.
\]
For the $V_R$ case we find
\[  M_{eff} \sim \frac{M_{GUT}}{\epsilon} \,.
\]
For a supersymmetry breaking scale $M_S$ in the TeV region, we conclude that the lifetime of the proton
is consistent with the experimental bounds  for an effective mass $M_{eff}$  a few times larger  than $M_{GUT}$ which can be satisfied
 for~\footnote{Since the mass insertion  $\langle \Phi\rangle 5_H\bar 5_{\bar x} \propto v_2$ one would expect that
   for $v_2\ll 1$ the contribution of the graph to Proton Decay would be small.  However,  the element $b$
 cancels the effect because it is also proportional to $b\propto v_2$.}
 $v_2>v_1$ and $\eps<1$.

There  are implications for the  $\mu$ term given by   $\mu\sim  v_1v_2v_3$ (see Eq.~\ref{Muterm}).
Since $v_1, v_2$ cannot be small, in order to sufficiently suppress this term
we must have  $v_3=\langle \theta_{43}\rangle/M_{GUT}\ll 1$.  On the other hand, the smallness of $v_3$ suppresses
also the mass scale of the lighter generations and  might lead to inconsistences with the experimental values.
We should recall, however, that there are significant contributions to the fermion masses from  one-loop
gluino exchange diagrams~\cite{Banks:1987iu} implying masses of the order
 {\small $m_{u/d}\propto\frac{a_3}{4\pi}A_q\frac{m_{\tilde q}m_{t/b}}{m_{\tilde g}^2}$}
for the up/down quarks, where $A_q, m_{\tilde g}, m_{\tilde q}$ are respectively the trilinear parameter, the  gaugino and squark masses.

We have already stressed that  the presence of additional vector-like 5-plets in the  model under consideration,
is compatible with a smaller value of the unified gauge coupling $g_U\sim 10^{-1}$ at the GUT scale.
This has significant implications for the proton  decay rate which occurs through the exchange of the gauge bosons
(dimension six operators).  For the well known case  $p\to e^+\pi^0$ the life-time is estimated to
be~\cite{Hisano:2013exa,Murayama:2001ur,Dutta:2016jqn}
\be
\tau(p\to e^+\pi^0)\approx  8\times 10^{34}{\rm years}\;\times
\left( \frac{a_{U}^0}{a_U}\times \frac{2.5}{A_R} \times\frac{0.015 {\rm GeV}^3}{a_H}\right)^2
     \times  \left(\frac{M_V}{10^{16}{\rm GeV}}\right)^4\,\cdot
     \ee
The various quantities in the above formula are as follows:  $a_U^0=\frac{1}{25}$ is assumed to be the value
of the unified gauge coupling in the minimal $SU(5)$,  while $a_U$ stands for its  value for the present model which
is taken to be $a_U\approx \frac{1}{8.5}$ (see  section 5). The factor $A_R$ takes into account the various
renomalisation effects, $a_H$ is the hadronic matrix element and $M_V$ denotes the mass of the gauge boson mediating
the process $p\to e^+\pi^0$.  Comparing with the recent experimental limit~\cite{Miura:2016krn}
$\tau(p\to e^+\pi^0)\ge 1.6\times 10^{34}$ years, we find a lower bound on the mass of the gauge boson $M_V\ge 1.14\times 10^{16}$ GeV, which is reciting since it just below the GUT scale predicted in this model $M_U\approx 2.04\times 10^{16}$ GeV.

\subsection{Variation with new physics predictions accessible at LHC }

In this section we consider the possibility of predicting new physics phenomena (such as diphoton events)
from relatively light ($\sim$ TeV) scalars and triplets.  The model discussed so far cannot accommodate a
process such as the diphoton event, since there is no direct coupling $ D^cD S$ with a light singlet $S$.
Indeed, the only singlet coupled to  $D^c,D$ is $\theta_{14}$ which acquires a large vev and decouples.
To circumvent  this we briefly present a modification of the above model by assuming the following non-zero vevs,
\be
 v_1=\langle \theta_{13}\rangle ,\; v_2=\langle \theta_{34}\rangle ,\; v_3=\langle \theta_{43}\rangle\,,  \label{newvevs}
\ee
and we maintain the same assignments for the fermion generations listed in Table~\ref{modA}.
The mass matrices for the up, down quarks and charged leptons, are given by
\ba
m_u\sim
\left(\begin{array}{ccc}
v_1^2&v_1^2v_2&v_1\\
v_1^2v_2&v_1^2v_2^2&v_1v_2\\
v_1&v_1v_2&1
\end{array}
\right)h_t\langle H_u\rangle\,,\;
m_{d,\ell}\sim
\left(\begin{array}{ccc}
v_{1}^2&v_{1}v_{3}&v_{1}\\
v_{1}^2v_{2}&v_{1}&v_1v_2\\
v_{1}&v_{3}&1
\end{array}
\right)h_b\langle H_d\rangle,
\label{updownlepton}
\ea
where, as before, we have suppressed the Yukawa couplings expected to be of ${\cal O}(1)$.
We observe that the matrices exhibit the expected hierarchical
structure. Assuming a natural range of the vevs and Yukawa couplings
we  estimate that the fermion mass patterns are consistent with the observed mass spectrum.

\noindent
With this modification, the singlet  $\theta_{14}$ is not required to acquire a large vev and
it can remain as a light singlet $\theta_{14}=S'$. Through its superpotential coupling
\[   \theta_{14} \bar 5_{\bar x}5_x \to S' ( {D''}^c\,D+ H_u'H_d')\,,\]
where  ${D''}^c$ stands for  the linear combination ${D''}^c = \cos\phi D^c+ \sin\phi{\tiny } {D'}^c$,
$S'$ could contribute to diphoton emission.

\section{Summary} % and Conclusions}

F-theory appears to be a natural and promising  framework for constructing unified theories
with predictive power. The  SU(5) GUT model in particular, appears to be the most economic
unified group containing all those necessary ingredients to accommodate vectorlike fermions
that might show up in  future experiments. Therefore, in the light of possible
  new physics at the LHC experiments, in this letter, we reconsidered a class of F-theory $SU(5)$ models
aiming to concentrate on the specific predictions  and low energy  implications.

 \noindent
In the F-theory framework, after the $SU(5)$ breaking down to the Standard Model gauge symmetry, we end up with the MSSM
chiral mass spectrum, the Higgs doublet fields and usually a number of vector-like  exotics as well
as neutral singlet fields. We point out that we dispense with the use of large Higgs representations
for   the  $SU(5)$ symmetry breaking since the latter takes place by implementing  the mechanism
of the hypercharge flux. The corresponding $U(1)_Y$ gauge field remains massless by requiring the hypercharge
flux to be globally trivial. As a result of these requirements, the spectrum of the effective theory
and the additional abelian symmetries accompanying the GUT group, are subject to certain constraints.
In addition to the $SU(5)$ GUT group, the  model is subject to additional symmetry restrictions emanating
from the  perpendicular `spectral cover' $SU(5)_{\perp}$ group, which in the effective theory reduces
down to abelian factors according to the `breaking' chain
\[ SU(5)_{\perp} \supset U(1)_{\perp}^4  \myeqr  U(1)_{\perp}^3 \]
where $Z_2$ is the monodromy action, chosen for this particular class of models under discussion~\footnote{For
the $SU(5)_{\perp} $ spectral cover symmetry, the possible monodromies fall into a discrete subgroup
of the Weyl group $W(SU(5)_{\perp})\sim S_5$, with $S_5$ being the permutation symmetry of five objects.}.
A suitable choice of fluxes along these additional abelian factors is responsible for the chirallity of the
$SU(5)$ GUT representations and their propagation on the specific matter curves presented in this paper.

\noindent
 In practice,  the effects of the remaining spectral cover symmetry  in the low energy effective theory
 are described by a few integers (associated with fluxes) and the `charges'-roots $t_i, i=1,2,\dots, 5$
 of the  spectral cover fifth-degree polynomial
  where two of them, namely $t_{1,2}$, are identified under the action of the  monodromy
  $Z_2: t_1\leftrightarrow t_2$ applied in this work.

\noindent
 The implementation of the hyperflux symmetry breaking mechanism has additional interesting
effects. As is known, chiral matter and Higgs fields reside on the intersections (i.e., Riemann surfaces, dubbed
here as matter curves and characterised by the remaining $U(1)$ factors through the `charges' $t_i$)
of seven branes  with those wrapping the $SU(5)$ singularity. In general the various
intersections are characterised by distinct geometric properties and  as a consequence flux restricts
differently  on each of them, while implying splittings of the $SU(5)$ representation content in certain
cases.  As a result, in the present model doublet Higgs fields  are accommodated
on matter curves which split the $SU(5)$ representations realising an effective doublet-triplet splitting
mechanism in a natural manner. More precisely,  this ammounts to removing one triplet
 from the initial Higgs curve with the simultaneous appearance (excess) of another one on a different matter curve.
This displacement however is enough to  allow a light mass term for the Higgs doublets
while heavy  triplet-antitriplet mass terms originate from different terms leading to
suppression of baryon number violating processes.
Chiral fermion generations are chosen to be accommodated on different
matter curves, so that a Froggatt-Nielsen type mechanism is implemented to generate the required hierarchy.
Furthermore, certain Kaluza-Klein modes are associated with the right-handed neutrino fields implementing
the  see-saw mechanism through appropriate mass terms with their left-handed counterparts.

\noindent
The additional spectrum in the present model consists of neutral singlet fields
as well as  colour triplets and Higgs-like doublets comprising
complete $SU(5)$ vector-like pairs in $5+\bar 5$ multiplets, characterised by non-trivial $t_i$-`charges'.
Some singlet fields  are allowed to acquire vevs at the TeV scale inducing  masses of the same order for
the vector-like exotics through the superpotential terms.
Such `light'  exotics contribute to the formation of  resonances producing excess
of diphoton events which could be discovered in future LHC experiments.  A RGE analysis shows that
the resulting spectrum is consistent with gauge coupling unification and the predictions of the
third family Yukawa couplings.

\vspace{1cm}
{\bf Acknowledgements}. {\it The authors would like to thank R. Blumenhagen, S.F. King  and Tianjun Li for reading
	 the manuscript 	and helpful comments.
	G.K.L. would like to thank the Physics and Astronomy Department and Bartol Research
Institute of the University of Delaware for kind hospitality. Q.S. is supported in part by the DOE grant ``DOE-SC-0013880''.}

\end{document}